# A Dynamic Computational Model of Head Sway Responses in Human Upright Stance Postural Control During Support Surface Tilt

Vittorio Lippi[1,2][a], Christoph Maurer[2][b] and Stefan Kammermeier[3][c]
[1]*Institute of Digitalization in medicine, Faculty of Medicine and Medical Center, University of Freiburg, Freiburg im Breisgau, Germany*
[2]*Clinic of Neurology and Neurophysiology, Medical Centre-University of Freiburg, Faculty of Medicine, University of Freiburg, Breisacher Straße 64, 79106, Freiburg im Breisgau, Germany*
[3]*Klinikum der Universität München, Ludwig-Maximilians-Universität LMU, Neurologische Klinik und Poliklinik, Marchioninistraße 15, 81377 München, Germany*
{Vittorio.Lippi, Christoph.Maurer}@uniklinik-freiburg.de, stefan.kammermeier@med.uni-muenchen.de

Abstract: Human and humanoid posture control models usually rely on single or multiple degrees of freedom inverted pendulum representation of upright stance associated with a feedback controller. In models typically focused on the action between ankles, hips, and knees, the control of head position is often neglected, and the head is considered one with the upper body. However, two of the three main contributors to the human motion sensorium reside in the head: the vestibular and the visual system. As the third contributor, the proprioceptive system is distributed throughout the body. In human neurodegenerative brain diseases of motor control, like Progressive Supranuclear Palsy PSP and Idiopathic Parkinson's Disease IPD, clinical studies have demonstrated the importance of head motion deficits. This work specifically addresses the control of the head during a perturbed upright stance. A control model for the neck is proposed following the hypothesis of a modular posture control from previous studies. Data from human experiments are used to fit the model and retrieve sets of parameters representative of the behavior obtained in different conditions. The result of the analysis is twofold: validate the model and its underlying hypothesis and provide a system to assess the differences in posture control that can be used to identify the differences between healthy subjects and patients with different conditions. Implications for clinical pathology and application in humanoid and assistive robotics are discussed.

## 1 INTRODUCTION

### 1.1 Overview

The analysis of body segment sway during a perturbed upright stance is an established experimental method to investigate posture control and the underlying neurological processes in human subjects (e.g., Assländer & Peterka, 2016; Goodworth & Peterka, 2018) and to test the balance capabilities of humanoid robots (Pasma et al., 2018; Zebenay et al., 2015). Perturbation of upright stance allows insights into the internal control mechanisms' limitations and pinpoint the cause of possible deficits. These alterations may be externally applied or initiated by voluntary motion of the subject itself.

For the underlying data of our proposed model, upright standing subjects were studied during the passively applied small-angle rotational disturbance with multiple superimposed sinusoidal frequencies of their support surface, where the center of rotation was placed at the ankle joint. Three groups of subjects were investigated: healthy control subjects and neurodegenerative disorders with movement control impairment: Progressive Supranuclear Palsy PSP and Idiopathic Parkinson's Disease IPD. The body segment movements were recorded with 3D ultrasound motion capture detailed in Methods and Kammermeier et al., 2018. The proposed model fits data from human experiments in different conditions, i.e., eyes-open, and eyes-closed, combined with two amplitudes. The model was implemented with and without including nonlinearities observed in previous experiments to test how they capture the nonlinear responses typical of humans. Usually, larger disturbances are, in proportion, compensated more than smaller ones (Hettich et al., 2013 and 2015).

[a] https://orcid.org/0000-0001-5520-8974
[b] https://orcid.org/0000-0001-9050-279X
[c] https://orcid.org/0000-0003-0158-889X

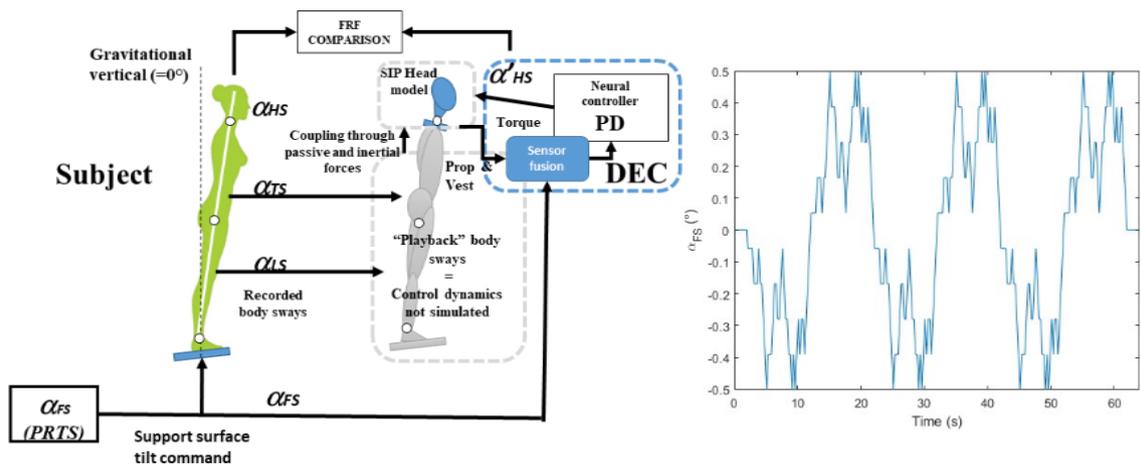

Figure 1: General Scheme of the simulated system. The recorded sways for the legs and trunk and the known PRTS input profile are used to simulate the neck movement. On the left, the profile of the PRTS stimulus $\alpha_{FS}$ with peak-to-peak amplitude 1°.

Based on these results, a model for a quantitative description of the posture control task in terms of input (perturbation stimulus) and output (body segment sway) was constructed, focusing on the head segment. The response can be represented as a transfer function in the frequency domain and developed further as a dynamic system of response regulation. The inherently unstable human upright posture requires dynamic feedback control based on sensory input.

The degrees of freedom vary depending on the scenario: a single-inverted-pendulum model (SIP), representing the torque provided around the ankles, is suitable to describe small movements produced by small perturbations. An added hip segment with intersegmental coordination (Hettich et al., 2014) results in a double-inverted pendulum model (DIP). Involving the knees in the compensation of more intense and challenging disturbances with a third degree of freedom (Atkeson & Stephens 2007) suggests that responses could be possibly dictated by optimization criteria penalizing the amplitudes of joint angles excursions and the applied torques, at least in transient responses to impulsive disturbances.

More than 2 degrees of freedom are usually modeled when applying posture control models to humanoids in order to cover the number of actuators in their kinematics (e.g., three in Ott et al., 2016) or to model the interaction between a human and a wearable exoskeleton (e.g., four in Lippi & Mergner, 2020). In most cases, the control of the head is not taken into account, and the head is considered together with the rest of the upper body since the head is considered a minor fraction of the upper body's mass (about 16%, in De Leva, 1996).

This paper addresses the modeling of head control as a modular control system, associating each degree of freedom with a feedback controller (Hettich et al., 2014; Lippi & Mergner, 2017). The specific formulation of the control problem (§ 2.1) allows the identification of the control parameters just for the module controlling the neck without including a full model of the other joints.

The internal representation of human body motion in space is construed from three main sensory input qualities for the purpose of motor control (e.g., summarized in Dieterich & Brandt, 2019): vestibular angular acceleration and gravitational tilt from the inner ear; visual alignment to vertical and horizontal references from the eyes; and proprioceptive joint and muscle sensors for their mutual relative positions throughout the body. Each biological sensor type has optimal sensory operating ranges of kinetic motion and stationary position to overlap with the weak ranges of other sensors. Two of these three sensory contributors reside in the freely mobile human head, unlike in current self-propelled humanoid robots (e.g., Boston Dynamics Atlas). Positioning vision and vestibular system in a small mobile mass on top instead of within the main mass of the body has several practical and evolutionary advantages. This feature could not yet be practically implemented in humanoid robotics by abundant technical and computational constraints.

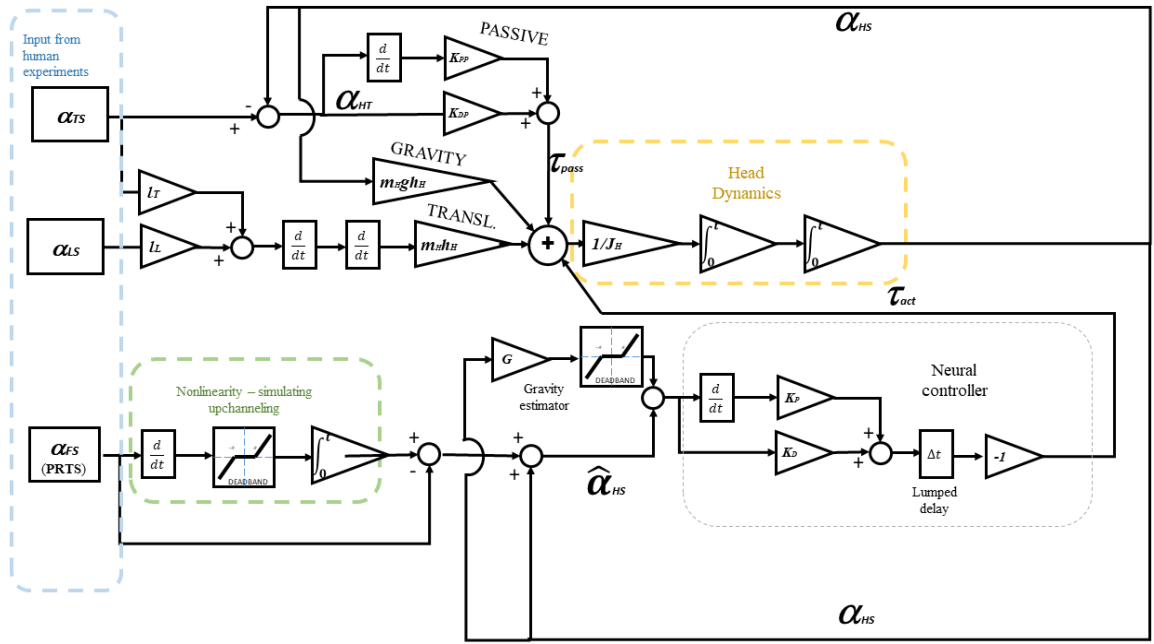

Figure 2: The Dynamic system from equations (1) to (7). Here the controlled variable is the head-in-space angle $\alpha_{HS}$. To control the head-to-trunk angle $\alpha_{HT}$, such a variable should be provided as input for the neural controller instead of $\alpha_{HS}$. As the reference is assumed to be 0° the controlled variable equals the error $\varepsilon$ in (1). The recorded body sway for the trunk and legs is used as input. In contrast, the recorded head sway is used outside the simulation to optimize the parameters and evaluate the result.

Therefore, understanding this sensorimotor integration in healthy humans, in comparison to human disease conditions, which particularly impair the integration of head-based sensory information in motor control, may be considered essential to future developments in several fields of science:

- A model that can emulate healthy and degenerative brain diseases may help to understand analogies in neuroanatomical and physiological networks so that tools for diagnosis, progression monitoring, and effect graduation of possible targeted therapies for specific aspects of the degenerating neural network might be developed.
- For humanoid robotics, a functional model with known weak points similar to human disease conditions may help to develop a robust computational framework to integrate sensor arrays remote to the main robot body, analog to the human head. This may indirectly open the main robotic body space for other systems, like advanced propulsion or increased payload.

## 1.2 State of the Art

The modeling of posture control as a dynamic system is widespread in the analysis of human experiments and humanoid robotics. There is no consent about how the "actual human control system" is, and various approaches are proposed in the literature, depending on the analyzed task and the question under investigation. Usually, models describe the system as an inverted pendulum (single or multiple) where the joint torques are controlled (Hettich et al., 2014; Maurer et al., 2006; Thomas Mergner, 2010; Peterka, 2002); other models include muscle activity (Souza et al., 2022) and tendon dynamics (Loram et al., 2004, 2005b, 2005a; P. G. Morasso & Schieppati, 1999) specifically. The effort in designing the controller can be oriented toward representing the biomechanics (Alexandrov et al., 2017, 2015), investigating some neurological processes such as sensor fusion (Thomas Mergner, 2010; Peterka, 2002), or a more general model of human neural control (Jafari et al., 2019; McNeal & Hunt, 2023), or based on some optimization (Atkeson & Stephens, 2007; Jafari & Gustafsson, 2023; Kuo, 2005). While

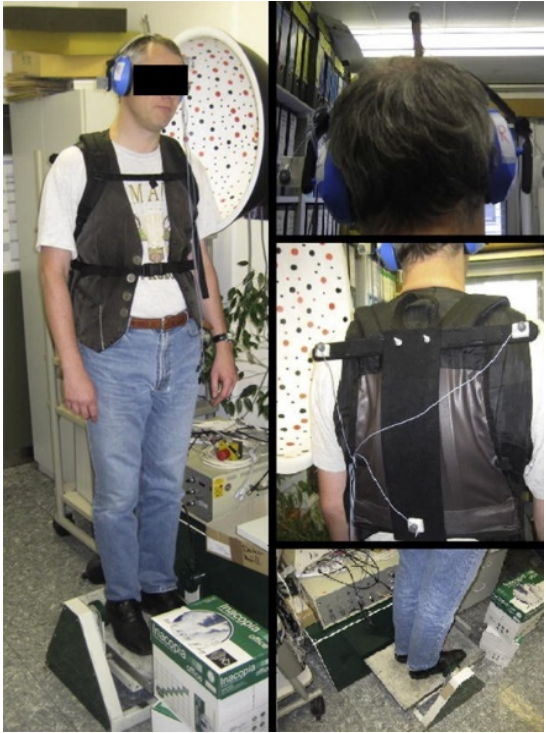

Pictures from Kammermeier et al. (2018).

Figure 3: The experimental setup: the test subject stands on a tilting platform. The subject's feet were placed within marked positions, with the heels together and the tips spread 15° apart, while the arms hung loosely by the sides.

the optimization provides an interesting insight into why the subjects may behave in a certain way and, by definition, optimized controls for humanoids, the optimal control is rarely totally human-like as human behavior is reasonably the result of a trade-off between typical control system requirements as tracking precision and energy efficiency and biological peculiarities that are less obvious from the engineering point of view such as reducing the effort of the nervous system or dealing with the fear of falling.

In general, the more a task is complicated, the more degrees of freedom are involved; for example, quiet stance can be modeled with a single inverted pendulum where the body sway is controlled by ankle torque (P. Morasso et al., 2019) at least to predict center of mass (COM) sway. However, there is a small activity of other joints (Pinter et al., 2008). Applying optimization to the compensation of disturbances (push) of different amplitudes reveals the emergence of strategies involving more joints, going for ankle strategies for small pushes to knee bend for larger ones, passing through hip-ankle strategies for stimuli in between (Atkeson & Stephens, 2007). As anticipated above, the control of the neck is not common in literature, as usually, the upper body is considered altogether and referred to as HAT (head, arms, and trunk). In (Kilby et al., 2015), kinematic 3D models with different numbers of degrees of freedom (ranging from 1 to 7) for posture control are proposed for the analysis of different tasks recorded with subjects, i.e., quiet standing, standing on foam, and standing on one leg. The simple model has just the ankle joint, while the most complex has ankles, knees, hips, and neck. No control system is proposed, and the analysis is performed on the variance of joint angles. The results show that the model with the neck (7 DOF) is the most accurate in fitting the sway of the center of mass and confirm that with increasingly difficult tasks (standing on foam is more demanding than quiet standing on a firm surface, and standing on one leg is even more difficult). This suggests that including the neck in the control system can be beneficial in understanding human posture control in the general case.

In this paper, the focus is on the neck with the twofold objective of, on the one hand, exploring how the DEC model can be extended with an additional degree of freedom and testing whether the modularity of the DEC control can be applied to study local characteristics of the control, e.g., in this case, the stiffness of the neck, and, on the other hand, provide a tool for the specific analysis of PSP and IPD patients.

## 2 MATERIALS AND METHODS

### 2.1 Clinical Experimental Data

The proposed model emulates the 3D motion analysis data obtained from three groups of human subjects. The overall setup is described in detail in Kammermeier et al. 2018; here, 19 healthy elderly subjects are considered. All participants gave written informed consent, and data was anonymized at study inclusion following the Helsinki Declaration and the local ethics committee (142/04; Ethikkommission der Medizinischen Fakultät).

All subjects were placed on a remotely controllable platform (Toennis) as shown in Fig. 3, allowing a front-to-back tilt through the ankle joint axis. The stimulation paradigm involved a resting condition and small-angle (0.5° and 1°) rotational disturbances with multiple superimposed sinusoidal frequencies (0.05-2.2Hz) for 60 seconds; each maximum angular displacement was tested in eyes open EO and eyes closed EC conditions.

3D motion capture was performed with ultrasound receivers (Zebris 3D) and re-sampled to 100Hz. The sensors placement is shown in Fig. 4 The final dataset included the platform motion track and the position signals of three markers for the head, three for the chest, one for the lower spine, and one each at the knees along the time domain.

## 2.2 Posture Control Model

**The Control Model** is based on the *disturbance identification and compensation* (DEC) principle (Thomas Mergner, 2010) and its implementation as a modular control system (Hettich et al., 2014; Lippi et al., 2016; Lippi & Mergner, 2017). The DEC is based on the hypothesis of a servo controller for body position (Merton, 1953), complemented with the estimation and compensation of external disturbances based on sensory input (T Mergner et al., 1997; Thomas Mergner & Rosemeier, 1998). The servo controller is implemented as a PD, *proportional derivative* controller. The compensation, which generally includes support surface tilt and acceleration, gravity, and external push, here is considered only for gravity and support surface tilt. Such compensation is implemented in a *Störgrößenaufschaltung* fashion, literally meaning *disturbance control* in German, i.e., a feed-forward compensation of disturbances based on sensory input that allows us to estimate the disturbance itself (Bleisteiner et al., 1961). The DEC controller is mainly used to predict steady-state responses, unlike other models oriented to transient responses (e.g., Allum & Honegger, 1992; Küng et al., 2009).

**The Control Equations** regulating the action of the servo controller for the neck are:

Active Torque
$$\tau_{act} = (K_P + K_D \frac{d}{dt})(\varepsilon + \hat{G}) \quad (1)$$

Passive Torque
$$\tau_{Pass} = (K_{PP} + K_{PD}\frac{d}{dt})(\alpha_{HT}) \quad (2)$$

Gravity Compensation
$$\hat{G} = K_G \vartheta_G(\alpha_{HS}) \quad (3)$$

Support Surface Tilt Compensation
$$\hat{\alpha}_{HS} = \alpha_{HS} + \alpha_{FS} - \hat{\alpha}_{FS} \quad (4)$$

Where $K_P$ and $K_D$ are the coefficients of the PD controller, $\varepsilon$ the error on the controlled variable that can be the estimated head in space position $\hat{\alpha}_{HS}$ from (4) or $\alpha_{HT}$, the angle between head and trunk, and $\hat{G}$ is the estimated gravity torque from (3). Expressing the disturbance as an additional input for the PD exploits the derivative as an anticipation effect. In robotics applications, the compensation can have its own PD parameter to allow for fine control (Ott et al., 2016), here only a PD according to previous works where the DEC is used to model human responses (e.g., Georg Hettich et al., 2014; Thomas Mergner et al., 2009). $K_{PP}$ and $K_{PD}$ are the passive stiffness and damping associated with the neck. $\alpha_{HS}$ in (3) and (4) is the head-in-space angle (with respect to the gravitational vertical) that here is assumed to come from the vestibular system without any modeled noise. $K_G$ is a coefficient associated with gravity (that usually, in humans, is slightly under-compensated, taking into account the additional torque produced by the servo loop). The function $\vartheta_G()$ is a dead-zone nonlinearity defined as

$$\vartheta_G(\alpha) = \begin{cases} \alpha + \theta_G & \alpha < -\theta_G \\ 0 & -\theta_G < \alpha < \theta_G \\ \alpha - \theta_G & \alpha > \theta_G \end{cases} \quad (5)$$

Where the threshold $\theta_G$ is a non-negative parameter, the estimation of support surface tilt is affected by a nonlinearity also reflected in the estimated $\hat{\alpha}_{HS}$ in (4). Specifically, the foot-in-space estimation is performed by fusing the vestibular angular velocity signal with the proprioception of all the joints from the head to the ankle, a nonlinear function $\vartheta_{FS}()$ with the same formal expression of (5) and another threshold $\theta_{FS}$ is then applied to the resulting velocity signal. Here, as the proprioception and vestibular signals are modeled as ideal, the $\alpha_{FS}$ known from the experiment design (§2.2.) is used to produce the following estimation

$$\hat{\alpha}_{FS} = \int_0^{t_1} \vartheta_{FS}(\frac{d}{dt}\alpha_{FS}) d\tau \quad (6)$$

Where $t_1$ is the current time and the initial condition of the estimator is assumed to be $\hat{\alpha}_{FS} = 0$. Again for the assumption of ideal proprioceptive signals, the error $\alpha_{FS} - \hat{\alpha}_{FS}$ is propagated directly to $\hat{\alpha}_{HS}$, leading to eq. (4). The nonlinearity $\vartheta_{FS}()$ explains the fact observed in subjects that smaller stimuli are associated with larger gains in the responses (Hettich et al., 2011). Although introducing an error in the tracking of body sway that prevents asymptotic stability, the dead band does not make the system unstable; in fact, it can be demonstrated that the system is Lyapunov stable (Lippi & Molinari, 2020).

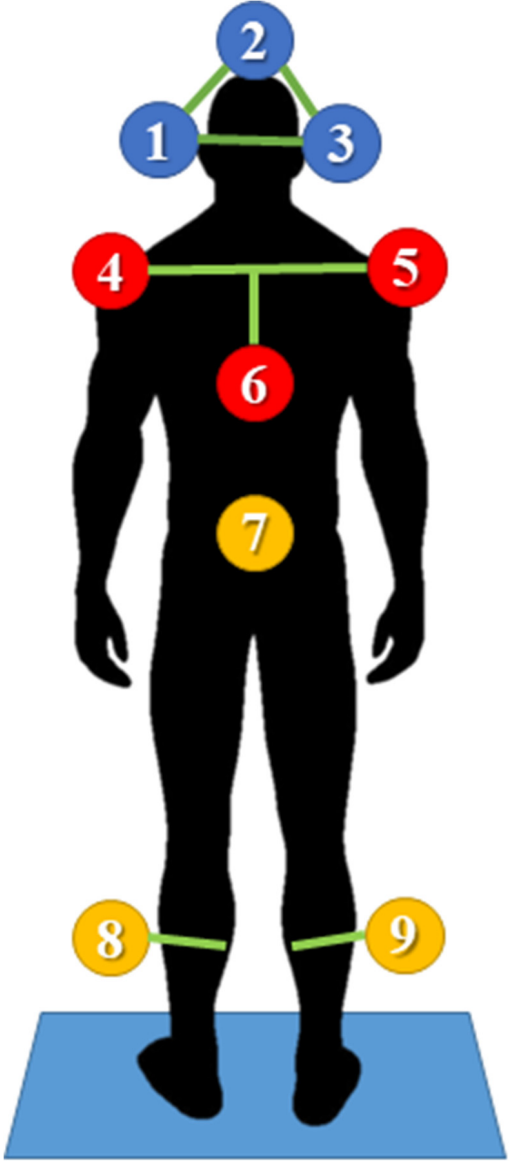

Figure 4: 3D ultrasound position markers (Zebris system) placement: head (1,2,3), upper chest (4,5,6), hip (7) and lateral femoral epicondyles (8,9).

A lumped delay $\Delta t$ representing all the delays in the loop affects the active control (it is in series with the PD). The sources of the delay are the sensory input and the motor control. For the control of the ankle, they are usually estimated to be between 80 and 200 ms, depending on the subject and the test conditions (Antritter et al., 2014; Li et al., 2012; Molnar et al., 2021). In general, the delay of peripheral body joints is expected to be larger than the one associated with joints closer to the brain, e.g., the delay in the control loop of the hip is smaller than the one of the ankle (e.g., 70 ms and 180 ms respectively in G Hettich et al., 2011). This suggests that the delay in the neck control loop will be smaller (as confirmed in the Results §3). Delay is an important parameter as it is observed that an added delay of about 70 – 80 ms can already be perceived (Morice et al., 2008) and can degrade the performance in tracking tasks (Lippi et al., 2010) and; hence it is reasonably one of the parameters characterizing the performance also in posture control as suggested by (Li et al., 2012; Lockhart & Ting, 2007; Molnar et al., 2021; Van Der Kooij et al., 1999; Vette et al., 2010). Furthermore, delay appears to be a determining factor in the different performances between young and elderly subjects (Davidson et al., 2011; Qu et al., 2009).

A schema of the simulated system with an overview of the DEC control, including the simulated dynamic system, is shown in Fig. 1.

**The Dynamic of the Head** is simulated as a single inverted pendulum (SIP) characterized by the weight and the moment of inertia of the head on which the active toques $\tau_{act}$ and $\tau_{Pass}$ from (1) and (2). The translation produces a further effect due to the sway of the legs and the trunk resulting in the following dynamic system, where the small angle approximations $\sin\alpha \approx \alpha$ and $\cos\alpha = 1$ are applied:

$$\begin{cases} \dot{\alpha}_{HS} = \ddot{\alpha}_{HS} \\ \ddot{\alpha}_{HS} = (\tau_{act} + \tau_{pass} + G + T_{acc})/J_H \\ G = m_H g h_H \alpha_{HS} \\ T_{acc} = (\ddot{\alpha}_{LS} l_L + \ddot{\alpha}_{TS} l_T) h_H m_H \end{cases} \quad (7)$$

where $J_H$ is the head moment of inertia, $m_H$ is the head mass, $h$ e is the height of the head center of mass, and $g = 9.81\ m/s^2$ is the gravity acceleration constant. $l_L$ and $l_T$ are the lengths of the trunk and leg segments, respectively. A standard set of anthropometric parameters (De Leva, 1996; Winter, 2009) is used in all the simulations with no specific adaptation to the single subject. Anthropometric parameters are reported in Table 1. The full dynamic system is shown in detail in Fig. 2.

## 2.3 The Dataset

**The Frequency Response Function, FRF**, is an empirical transfer function between the stimulus and the body sway. The input support surface tilt has a Pseudo Random Ternary Sequence (PRTS) profile (Peterka, 2002). The fact that the system is pseudo-random prevents the subjects from using prediction, as it is expected that the user can learn the sequence (Thomas Mergner, 2010). However, a study (Assländer et al., 2020) suggests that the adaptation observed in subjects after repeating the experiments may be due to learning the dynamics of the body in the experimental scenario, as they occur both with pseudo-random and rhythmic stimuli. Robotic experiments showed how DEC could improve performance when integrating an online learning system to predict future disturbances (Lippi, 2018). In the presented model, no prediction or learning will be taken into account. The peak-to-peak amplitudes used in the presented experiments are 0.5° and 1°. The PRTS power spectrum has a profile with peaks separated by ranges of frequencies with no power (Joseph Jilk et al., 2014; Lippi et al., 2020; Peterka, 2002). The response is evaluated for specific frequencies where the PRTS spectrum has peaks (see Fig. 5, top). The FRF is computed with the Fourier transform of the input U and the output Y as

Table 1: Anthropometric parameters used in all the simulations.

| Parameter | Symbol | Value |
|---|---|---|
| Head moment of inertia | $J_H$ | 0.4797 kg/m |
| Leg length (ankle to hip) | $l_L$ | 0.8543 m |
| Trunk length (hip to neck) | $l_T$ | 0.5011 m |
| Head COM height | $h_H$ | 0.2053 m |
| Head mass | $m_H$ | 4.5 kg |

where $G_{UY} = U * Y$ and $G_U = U * U$ are empirical estimations of the cross-power spectrum and the input power spectrum ($*$ is the element-wise product). The spectrum is then transformed into a vector of 11 components by averaging the FRF over neighboring ranges of frequencies, as illustrated in Fig. 5. The choice of the frequencies and their overlapping follows the method described in (Peterka, 2002), but here is adapted to the 11 frequencies considered as the PRTS used here is shorter in time (one cycle is 20 seconds, and the signal is repeated three, in contrast to the six cycles of 60.5 s of the original work). The averaging was initially proposed to obtain a visualization with the frequencies almost equally spaced on a logarithmic scale, typically used to plot the FRF. The final representation of the FRF is a function of the 11 frequencies

$f = [0.1, 0.3, 0.6, 0.8, 1.1, 1.4, 1.8, 2.2, 2.7, 3.5, 4.4] Hz$

Notice that a transfer function represents a linear system input-output relationship assuming no transient effect due to initial conditions. It is known that human postural responses are nonlinear (e.g., see *sensory reweighting* in Assländer & Peterka, 2016), so the FRF is intended as a representation of the output of a trial and not as a general model for the subject.

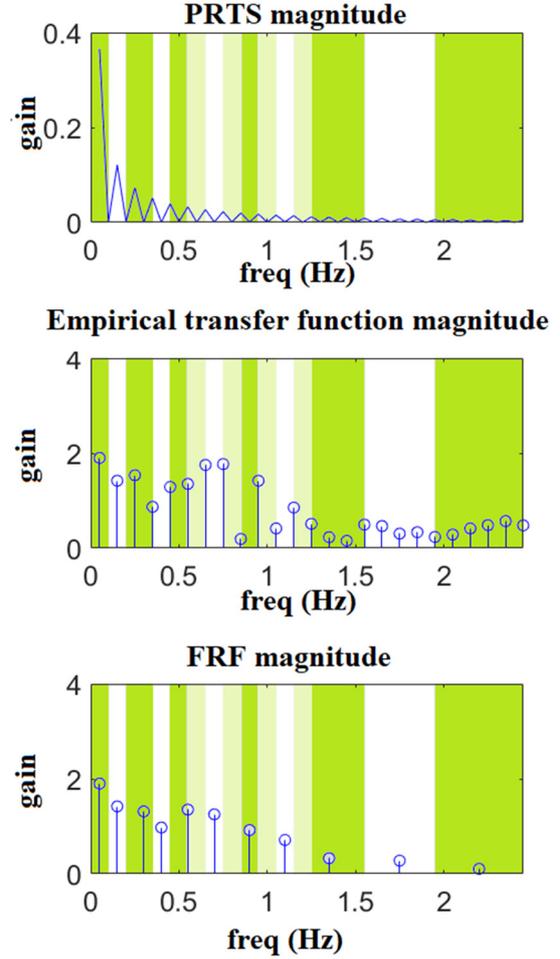

Figure 5: Spectrum and example of FRF. Above: The magnitude of the DFT of the PRTS. Center: empirical transfer function from (8). Below: FRF resulting from the averaging of frequency bands. The bands on the background show the frequency ranges over which the spectrum is averaged: white and dark green represent ranges associated with groups of frequencies. The sets of frequencies overlap, with light green bands belonging to both contiguous groups, and a sample on the transition between two bands belongs to both groups. As the FRF is averaged in the complex domain, the average shown in the plot is not the average of the magnitudes.

$$H = \frac{G_{UY}}{G_U} \quad (8)$$

## 2.4 Fitting the Model

The model fits the data using a numeric research algorithm (Lagarias et al., 1998) implemented by the

Matlab function fminsearch.

The objective function to be minimized is the difference between the FRF from the experimental trial and the one produced by the simulation. The threshold $\theta_{FS}$ is set to 0.065 rad/s, as Georg Hettich et al. (2014) reported. Some variations of the model have been tested, specifically: i. a linear model with no dead bands ($\theta_{FS} = 0$, and $\theta_G = 0$), ii. a model with no dead band on gravity ($\theta_{FS} = 0.065$, and $\theta_G = 0$), and iii. a model where the threshold $\theta_G$ was unknown a priori and considered a parameter to be selected to fit the model. The best model, in the sense of having the smallest fitting error on average, was ii. This may sound paradoxical as the model iii. includes the possibility of setting $\theta_G = 0$; hence, it is expected to perform at least equally to ii. The problem is that adding one free parameter can sometimes decrease the performance of fminsearch. Specifically, in this setup, the gain nonlinearity introduced by the dead bands may not be easy to distinguish from the effect of the control gains.

It should be noted that the linear model can reproduce the nonlinear response (compare the different gains associated with the two different amplitudes in Fig. 6) as the parameters are optimized for each trial separately, showing a *reweighting* in different conditions. The introduction (6) improves the average result. It leads to a smaller variance in the gains parameters suggesting that it can explain the nonlinear response "automatically," as shown more in detail in the next section.

In the following model ii. is discussed, and the identified parameters are $K_P$, $K_D$ from (1), $K_{PP}$, $K_{PD}$ from (2), $K_G$ from (3), and the lumped delay $\Delta t$.

## 3 RESULTS

The average and the standard deviation parameters for the different conditions, i.e., Eyes closed (EC), eyes open EO, and peak-to-peak amplitudes of 0.5° and 1°, are reported in Table 2. In the following, the error consists of the norm of the difference between the FRFs of the data and the simulated model. For this reason, it has no unit of measure being a ratio between angles as the FRFs.

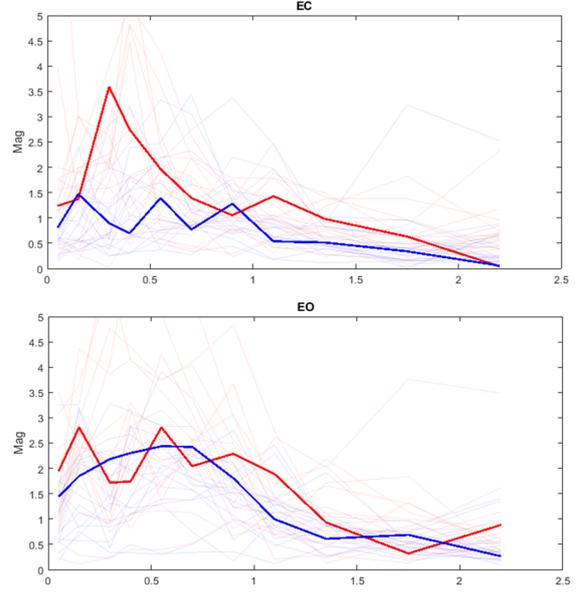

Figure 6: Average FRF magnitude for eyes closed (above) and eyes open (below) conditions. The red line is the response to 0.5°, and the blue line to 1°. The transparent lines on the background represent FRFs from single trials. The average is performed in a complex domain. The effect of the nonlinearity is evident with eyes closed: the gain is larger with the smaller stimulus that is under-compensated due to the effect of the thresholds as demonstrated in (Lippi & Molinari, 2020)

There is a significant difference in passive stiffness between the response to the stimuli of 0.5 and 1° amplitudes with EC ($K_{PP}$ was larger at 0.5° with p=0.04 tested with ANOVA). In contrast, the difference was smaller and not significant for EO (p=0.24). Slightly modified models have fit for comparison, and the detailed report of such fits goes beyond the space allowed for this work. The result can be summarized as follows: The linear model (with no dead-bands produces a larger error (2.7 compared to 2.1 of the final model) and a larger variance in the parameters (1.7) compared to the final model (1.4). A model where the proprioceptive variable (head-to-trunk angle $\alpha_{HT}$) was used instead of head-in-space $\alpha_{HS}$ as the control variable has produced worse results on average. The standard deviation of the proportional gains, both active ($K_P$) and passive ($K_{PP}$), is relatively large in the group of responses associated with 0.5°, especially for the EC case. This may be because when the oscillations of the trunk are small, the effect of those two gains is difficult to distinguish as $\alpha_{HT} \cong \alpha_{HS}$.

Table 2: Results of model fit for different visual conditions (EC, EO) and peak-to-peak (pp) amplitudes.

|  | **EC pp 0.5°** |  | **EC pp 1°** |  | **EO pp 0.5°** |  | **EO pp 1°** |  |
|---|---|---|---|---|---|---|---|---|
| *Parameter* | *mean* | *std* | *mean* | *Std* | *mean* | *std* | *mean* | *Std* |
| $K_P$ [Nm/rad] | 11.129 | 10.34 | 7.1597 | 1.1701 | 8.4829 | 5.2197 | 5.5033 | 1.7324 |
| $K_D$ [Nms/rad] | 1.5208 | 1.465 | 1.1701 | 0.91631 | 1.5034 | 0.85115 | 1.5213 | 0.8278 |
| $K_{PP}$ [Nm/rad] | 12.379 | 9.8733 | 6.9293 | 4.6557 | 13.583 | 11.079 | 10.139 | 5.3557 |
| $K_{PD}$ [Nms/rad] | 3.3978 | 3.1578 | 3.7765 | 3.0908 | 2.3834 | 1.1103 | 2.4437 | 0.85655 |
| $K_G$ | 1.049 | 0.49004 | 0.8516 | 0.56258 | 0.90084 | 0.45227 | 0.96701 | 0.36248 |
| $\Delta t$ [s] | 0.060489 | 0.040352 | 0.085444 | 0.061062 | 0.036357 | 0.028749 | 0.053899 | 0.017607 |

# 4 DISCUSSION CONCLUSIONS AND FUTURE WORK

The model that produced the best fit included the dead band nonlinearity from eq (6) but not the one on the gravity compensation. It is interesting to notice that the evident difference between EO and EC (see Fig. 6) is not reflected by a significant difference in the parameters in the neck control. This suggests that the differences in behavior between EC and EO are expressed in the sway of the trunk and the leg segments that affect the head as an input. The more relevant difference observed in parameters is the larger stiffness with small oscillations, suggesting that, with 0.5°, the head is moving with the trunk. These considerations suggest, on the one hand, that the DIP model is enough to predict the behavior of the upper body in healthy subjects (at least for the small oscillations considered). On the other hand, modeling the behavior of the neck may be beneficial for applications where head movements are involved explicitly, e.g., when the DEC is applied to an assistive device (Lippi & Mergner, 2020) if the device supports the head as well (e.g., an active version of Garosi et al., 2022). The patient's behavior is currently under investigation.

Models for perturbed posture control are used to explain the differences in body sway responses in terms of meaningful parameters, for example, sensory weighting in different conditions (Assländer & Peterka, 2014) or different subject characteristics such as stiffness or delay (A. Goodworth et al., 2023). In particular, the possibility to exploit the modularity of the DEC control to identify parameters relative to different joints will be used to identify differences between normal and pathological behavior expressed locally, e.g., see if the neck is stiffer in a group of patients.

The application to a bioinspired humanoid control system of the proposed neck model is straightforward: the up-channeled $\hat{\alpha}_{HS}$ value can be provided by the sensor fusion between the neck proprioceptive input (encoder) and the output of the module controlling the trunk $\hat{\alpha}_{TS}$ as specified in (Lippi & Mergner, 2017).

In this work, the fit of the model was performed with small amplitudes that are safe for patients (hence in the range of interest for future experiments). Future improvement in the model's validation can be achieved by integrating experiments with larger amplitudes. This may reveal nonlinear effects that are not visible with small head movements and, by producing larger trunk sways, allow for better differentiation between the effects of active and passive stiffness. For example, the gravity threshold that was not relevant in these experiments may become more visible.